\documentclass[12pt]{article}                                          
\usepackage{amssymb,amsmath,latexsym}
\usepackage{graphicx}
\usepackage[margin=1.0in]{geometry}
\usepackage{color}
\usepackage[pdftex,colorlinks=true,urlcolor=cyan,linkcolor=blue,citecolor=blue]{hyperref}

\title{Quasinormal mode spectra for odd parity perturbations in 
spacetimes with smeared matter sources}
\author {\tt Kumar Das $^{(a)}$\footnote{das.kumar582@gmail.com}, \hspace{4pt} Souvik Pramanik $^{(b)}$\footnote{souvick.in@gmail.com}
\hspace{4pt} Subir Ghosh $^{(a)}$\footnote{subir.ghosh20@gmail.com} \\ [10pt]
\small\em $^{(a)}$ S. N. Bose National Centre For Basic Sciences, 
JD Block, Sector III, Salt-Lake, \\ 
\small\em Kolkata-700098, India \\
\small\em $^{(b)}$ Physics and Applied Mathematics Unit, Indian Statistical Institute, 203 B.T. Road,\\ 
\small\em Kolkata - 700 018, India \\
\small\em $^{(c)}$ Department of Physical Science, Indian Institute of Science Education and Research Kolkata,\\
\small\em Mohanpur, India.
}

\date{}

\begin{document}

\maketitle
    \begin{abstract}
We have found the Quasi Normal Mode (QNM) frequencies of a class of static spherically symmetric spacetimes having a {\it {smeared}} matter distribution, parameterized by $\Theta$ -  an inherent length scale. 
Here our main focus is on the QNMs for the odd parity perturbation in this background geometry. 
The results  presented here for diffused mass distribution reveal  significant changes in the QNM spectrum.  This could be relevant for future generation (cosmological) observations, specifically to distinguish the signals of GW from a non-singular source in contrast to a singular geometry. We also provide numerical estimates  for the $\Theta$-corrected QNM spectrum applicable to typical globular cluster like spherical galaxies having a Gaussian spread in their mass distribution. We find that  the $\Theta$-correction to the GW signal due to  smeared distribution is  accessible to present day observational precision. 

\end{abstract}


\vspace*{0.07cm}

\newpage

\section{Introduction}
\label{introduc}

The recent detection of Gravitational Waves (GWs), from binary Black Hole (BH) mergers and Neutron stars by LIGO and Virgo collaboration \cite{Abbott:2016blz,Abbott:2016nmj,Abbott:2017vtc,TheLIGOScientific:2017qsa}, has provided us with a new window to study and understand physical processes at extreme conditions, where the role of gravity by far dominates  the other known forms of interactions in nature.  Since the metric to describe the gravitational collapse of a binary merger is unknown we fall back to  numerical relativity simulations from which we gain a fairly reasonable understanding of such realistic phenomena. For the present work we are assuming that these collapse events are generally consistent with the theoretical predictions of general relativity (GR) \cite{Will:2014kxa}, although later works \cite{kon} have shown that there are significant deviations. These observations now firmly suggest    studying the possibility of having alternative sources that can account for such deviations in  GW characteristic frequencies. Although an ultra-compact system of binaries is probably too  exotic a system for  producing strong GW signal to be detected at large distances, more common objects nevertheless can also produce GWs with frequencies that could be highly relevant for the next generation space-based GW detectors.

The popularly known GWs as observed by LIGO/Virgo collaboration are actually the Quasi Normal Modes or QNMs \cite{Berti:2005ys, Macedo:2016wgh,Kokkotas:1999bd, kon1}. The waveforms of these GW signals consist of three parts: 1) inspiral, 2) merger and 3) ring-down. The ring-down phase shows characteristic frequencies of oscillation corresponding to damped resonances of the remnant BH. These damped oscillations or QNMs encode information about the BH source. Applying the linear perturbation results, the ring-down portion of the signal may be used to discriminate between BHs and other possible sources. The damped modes in turn   possess a complex frequency whose real part corresponds to the oscillation frequency and whose imaginary part gives the lifetime. It is important to note that the QNM spectrum of a BH is completely characterized by the BH parameters, and does not depend on the initial conditions of the perturbations.

In this work our aim is to investigate the QNM frequencies for a spherically symmetric geometry having a smeared matter source. An interesting  approach was pioneered by Nicolini,  Smailagic and  Spallucci \cite{nicsp}. These authors introduced a (spherically symmetric) smeared source in the matter sector and solved the Einstein equation thereby obtaining a generalized form of Schwarzschild (black hole) metric that successfully cured the black hole singularity problem. It was also tentatively proposed to  identify the smearing scale with Planck length so that the metric can play a role in the context of quantum gravity. (For an exhaustive review see \cite{Nicolini:2008aj}.) Various aspects of this generalized black hole have been studied: its thermodynamics \cite{rabin},  effect of the smearing on AdS/CFT correspondence \cite{Pramanik:2015eka}, among others. As mentioned above,  the conventional Dirac delta source term for matter is replaced by a new type of matter source with the energy density  given by a Gaussian distribution function. The resulting geometry will be helpful to understand the dynamics of objects, which have approximately a Gaussian mass profile. From astrophysical point of view, such a Gaussian profile can be applied to study the dynamics of GWs for elliptical galaxies (\emph{e.g.} globular clusters having dense matter core in the center \cite{hogg:1965hg}). Besides that, this distribution is also relevant for the Dark matter profiles within galaxies (e.g. Press-Schechter mass distribution that is extensively used in the context of dark matter distribution profile \cite{Press:1973iz}). 
Therefore, the mathematical formulation of this study with smeared matter source will  be particularly interesting for astrophysical objects, where the length scales would be $\mathcal{O}(Ly)$ $( 1Ly\sim3\times 10^{-7} Mpc )$. In this paper, we can tentatively identify this length scale with  $\Theta$ - the smearing parameter.

Let us point out the proper perspective of our work in view of the recent works of Liang \cite{Liang:2018nmr, Liang:2018uyk} who has also made an exhaustive study of smearing effect on QNM. The results of Liang are valid up to $3$'rd order in WKB. On the other hand we have used the framework of \cite{Konoplya:2003ii} yielding results valid up to $6$'th order in WKB. We explicitly demonstrate that there are appreciable modifications when the latter are taken in to account.





 The organization of our paper is as follows --- in Sec.~\ref{methqnms}  we have briefly reviewed  the basic aspects of QNMs and an elementary method to obtain them for static spherically symmetric Schwarschild spacetime.  Sec.~\ref{gwgauss} deals with the gravitational perturbation of a spherically symmetric QG-inspired  spacetime. Here the analysis is made in four segments. 
In Sec~\ref{gwgauss1} we have computed the equations for the odd parity gravitational perturbation of this QG-inspired spacetime.
Then in Sec.~\ref{gwgauss2} and in Sec.~\ref{gwgauss3} we have obtained the QNM frequencies for this spacetime using the Ferrari-Mashoon formula \cite{Ferrari:1984zz} and also the WKB 6th order formula \cite{Konoplya:2003ii}. Here we discuss our results for the new QNMs by comparing them with the standard Schwarzschild QNMs and also with the QNMs for this QG-induced spactime, obtained earlier with 3rd order WKB method. Finally, in Sec.~\ref{gwgauss4} we discuss the relevance of the results from observational perspetives and in Sec.~\ref{conc} we conclude.  


\section{ QNM and determination of QNM spectrum: a brief review} 
\label{methqnms}

\noindent {\bf (a) Quasi-normal mode:}

\vspace{0.2cm}

 A black hole posses characteristic frequencies which arise from perturbations in it's spacetime geometry. 
Such perturbations of the BH geometry can originate in many different ways. For example a certain mass falling along the geodesic of the Schwarzschild spacetime can be considered as a perturbation on the background Schwarzschild geometry. In the presence of such distortion of the BH equilibrium, the BH system undergoes damped oscillations with complex frequencies. These frequencies are called quasi-normal modes (QNMs). Here the term `quasi' is referring to the fact that the frequencies are complex, thus they show  damping. While the conventional normal modes of compact classical linear oscillating systems are non-dissipative, for black hole QNMs \cite{Berti:2005ys}, the dissipations cannot be neglected, as the event horizon imposes necessary loss of energy.  
The real part of this QNM frequency corresponds to the oscillation frequency, whereas the imaginary part corresponds to the damping rate. From  astrophysical point of view, QNMs dominate an exponentially decaying ringdown phase at intermediate times in the GW signal from a perturbed BH \cite{Kokkotas:1999bd}. Moreover, they also govern the ringdown phase of gravitational systems produced by the merger of a pair of black holes \cite{Pretorius:2005gq,Campanelli:2005dd}. Since these QNMs are independent of the initial perturbation,  we can infer crucial information regarding the fingerprints (\emph{e.g} mass, charge and angular momentum of a BH \cite{Echeverria:1989hg}) of its source, the BH geometry.


Let us briefly discuss the equations governing the perturbation around a stationary spherically symmetric geometry \emph{i.e.} Schwarzschild spacetime. This spacetime is represented by the metric around a fixed spherically symmetric center-of-mass M
\begin{align}
ds^2 = -\bigg(1-\frac{2M}{r}\bigg)dt^2+\bigg(1-\frac{2M}{r} \bigg)^{-1}dr^2 + r^2 d\Omega^2
\label{schwarsc}
\end{align}
where, $d\Omega^2 = d\theta^2 +\sin^2\theta d\phi^2$. 

Now we consider a small non-spherical perturbation $h_{\mu\nu}$ such that the new perturbed metric is,
\begin{align}
 g_{\mu\nu} = \bar{\bf g}_{\mu\nu} + h_{\mu\nu}  \qquad\text{where,} \qquad \frac{|h_{\mu\nu}|}{|\bar{\bf g}_{\mu\nu}|}<<1 .
\end{align}
Here we denote the static generic background metric by $\bar{\bf g}_{\mu\nu}$. The inverse metric is then
\begin{align}
g^{\mu\nu} = \bar{\bf g}^{\mu\nu} - h^{\mu\nu} + \mathcal{O}(h^2).
\end{align}
The perturbed Christoffel symbols are given by
\begin{align} 
\Gamma^{\alpha}_{\mu\nu}& = \bar{\Gamma}^{\alpha}_{\mu\nu} + \frac 1 2 \bar{\bf g}^{\alpha\sigma} (h_{\sigma\nu,\mu}+h_{\sigma\mu,\nu}-h_{\mu\nu,\sigma} - 2h_{\sigma\kappa}\bar {\Gamma}^{\kappa}_{\mu\nu}) + \mathcal{O}(h^2) \nonumber \\
& \simeq \bar {\Gamma}^{\alpha}_{\mu\nu} + \delta\Gamma^{\alpha}_{\mu\nu},
\end{align}
where the $\bar{\bf \Gamma}$ is the Christoffel symbol for the unperturbed metric $\bar{\bf g}_{\mu\nu}$ and the small perturbation  $\delta\Gamma$'s  is given by,
\begin{align}
\delta\Gamma^{\alpha}_{\mu\nu} = \frac{\bar{\bf g}^{\alpha\beta}}{2}\big(\nabla_{\nu}h_{\mu\beta} + \nabla_{\mu}h_{\nu\beta} - 
\nabla_{\beta}h_{\mu\nu} \big).
\label{chris_eqn}
\end{align}


Using the definition of covariant derivative $\nabla_{\mu}$ (with $\nabla_{\mu}$ being with repect to $\bar{\bf g}_{\mu\nu}$) for the perturbed Christoffel symbol given in eqn.~\eqref{chris_eqn}, the vacuum Einstein field equation can be put into a more convenient form as
\begin{align}
\nabla_{\beta}\delta\Gamma_{\mu\nu}^{\beta} - \nabla_{\nu}\delta\Gamma_{\mu\beta}^{\beta} = 0.
\label{mod_eqn}
\end{align}
  Finally, putting the expression for $\delta\Gamma$ into eqn.~\eqref{mod_eqn} and employing gauge freedom, we get the second order differential equation for $h_{\mu\nu}$, 
\begin{align}
\Box h_{\mu\nu} - 2\bar{R}^{\rho}_{\,\sigma\mu\nu}h_{\rho}^{\,\sigma} = 0
\end{align}
in the TT (transverse traceless) gauge, where 
\begin{align}
\nabla^{\mu}h_{\mu\nu}=0 \quad \text{and} \quad h_{~\mu}^{\mu} = \bar{\bf g}^{\mu\nu}h_{\mu\nu} = h = 0.
\end{align} 

Now a generic perturbation, $h_{\mu\nu}$ of the spherically symmetric metric can be broken up into odd ($h_{\mu\nu}^{\text{odd}}$) and even ($h_{\mu\nu}^{\text{even}}$) parity components according to their transformation properties under parity \emph{i.e.} $(\theta,\phi)\to(\pi-\theta,\pi+\phi)$ \cite{Chandrasekhar:1985kt,Regge:1957td,Zerilli:1971wd,Rezzolla:2003ua}. 
Here, we will focus on the odd-parity perturbations $h_{\mu\nu}^{\text{odd}}$, also known as the axial perturbations. (We will comment about the even-parity perturbations towards the end.)
Its components are simplified by using the residual freedom to choose a proper gauge (\emph{e.g.} see \cite{Regge:1957td}) which eliminates all the highest derivatives in the angles $(\theta,\phi)$. Finally, the true gauge invariant axial perturbations are described by the functions $h_0(t,r)$ and $h_1 (t,r)$. 

The gravitational odd parity perturbations for this spherically symmetric spacetime are now described by the Regge-Wheeler equation 
\begin{align}
\frac{\partial^2Q(t,r)}{\partial t^2} - \frac{\partial^2Q(t,r)}{\partial r_{\star}^2} +  V_{\text{axial}}(r)Q(t,r) = 0
\label{swaschrl}
\end{align}
where $Q(t,r)$ is the gauge-invariant odd-parity variable, also known as Regge-Wheeler variable, and it is defined as
\begin{align}
Q(t,r)= \bigg(1-\frac{2M}{r} \bigg)\frac{h_1(t,r)}{r}
\end{align}
with $h_1(t,r)$ being an unknown function from the perturbation \footnote{The other component $h_0(t,r)$ can be removed by using the $\delta R_{\theta\phi}$ components of eqn.~\eqref{mod_eqn} ($\because \delta R_{\mu\nu} = \nabla_{\beta}\delta\Gamma_{\mu\nu}^{\beta} - \nabla_{\nu}\delta\Gamma_{\mu\beta}^{\beta} = 0$) \cite{Regge:1957td} } and the so called tortoise co-ordinate ($r_{\star}$) is defined as, 
\begin{align}
\frac{dr_{\star}}{dr} = \frac{1}{1- \frac{2M}{r}} .
\label{star_old}
\end{align}
Integrating eqn.~\eqref{star_old} one obtains for $r_{\star}$
\begin{align}
 r_{\star} &= r + 2M \ln\bigg({\frac{r}{2M}-1}\bigg) .
\end{align}
Since $r_{\star}\to \infty$ as $r\to\infty$ and $r_{\star}\to-\infty$ as $r\to2M$, so tortoise co-ordinate will be helpful in this context for it does not suffer from coordinate singularity near the event horizon at $r=2M$ (since $r_{\star}$ is pushed to $-\infty$ at horizon).  

Now extracting the time dependence in $ Q(t,r)$ as $Q(t,r) \sim e^{i\omega t} Q(r)$, eqn.~\eqref{swaschrl} takes the form
\begin{align}
\frac{\partial^2 Q(r)}{\partial r_{\star}^2} + \big( \omega^2 - V_{\text{axial}}(r)\big)Q(r) = 0 .
\label{schr}
\end{align}
The function $V_{\text{axial}}(r)$ is given by 
\begin{align}
V_{\text{axial}}(r) &= \bigg(1-\frac{2M}{r}\bigg)\bigg[\frac{l(l+1)}{r^2}-\frac{6M}{r^3} \bigg] .
\label{sc_po} 
\end{align}
The solutions of the eqn.~\eqref{schr} define the QNMs of the black hole with QNM mode frequencies $\omega$. Below we describe how to compute this frequency.

\vspace{0.4cm}

\noindent {\bf (b) Method for computing the QNM spectrum to $6$'th order in WKB approximation:}

\vspace{0.2cm}

There are various methods to determine the QNM spectrum of a black hole spacetime. Note that for Schwarzschild and Kerr black
holes, there exist the method of Leaver \cite{Leaver:1985ax}, who constructed exact eigensolutions of the radiative boundary-value problem of Chandrasekhar. Later Detweiler \cite{chandra} developed a stable numerical method
in order  to determine the
quasinormal frequencies with an arbitrary precision.  However, to the best of our knowledge, no such stable numerical method exists for the QG-inspired  spherically symmetric BH space-time \cite{Nicolini:2008aj},   that is capable of  evaluating the QNMs with arbitrary precision. Therefore, to find the QNMs we need to address the problem using approximations. One of the easiest semi-analytic way to determine the QNMs in frequency domain is the approximation of the effective potential by the P$\ddot{\text o}$schel-Teller potential. This method was suggested by Ferrari and Mashoon \cite{Ferrari:1984zz}.\footnote{For  details and review of the other methods see for example \cite{Berti:2005ys,Macedo:2016wgh}.} 

In this approach \cite{Ferrari:1984zz}, the main problem of finding the QNMs $\omega$ for $V^{\text{axial}}(r)$ is reduced to finding the bound state of an inverse potential given by the profile 
\begin{align}
V_{PT} (r_{\star})= \frac{V_0}{\cosh^2{\alpha(r_{\star}-\tilde{r}_{\star})}}. 
\label{pt_pt}
\end{align}
This is the P$\ddot{\text o}$schel-Teller potential, where  $\tilde{r}_{\star}$ is the point of extremum of the potential. Here $V_0=V_{PT}(\tilde{r}_{\star})$ is the height and $\alpha= \frac{1}{2V_0}\,\frac{d^2 V_{PT}}{dr_{\star}^2}\bigg|_{r_{\star}=\tilde{r}_{\star}} $ is the curvature
of the potential  at the extremum.
The bound state frequencies $\Omega(V_0,\alpha)$ of this potential are exactly known to be
\begin{align}
\Omega(V_0,\alpha) = \alpha \bigg[-\bigg(n+\frac 1 2 \bigg) + \bigg(\frac 1 4 + \frac{V_0}{\alpha^2} \bigg)^{1/2} \bigg] .
\label{bo_pt}
\end{align}
The proper QNM frequencies of the original potential in eqn.~\eqref{sc_po} are then obtained from eqn.~\eqref{bo_pt} by the parameter replacement $(V_0,\alpha) \to(V_0,i\alpha)$ \cite{Ferrari:1984zz}, 
and are given by the expression
\begin{align} 
\omega = \Omega(i\alpha) = \pm\sqrt{\bigg(V_0 - \frac{\alpha^2}{4} \bigg)} - i \alpha \bigg(n+\frac1 2 \bigg) .
\label{mash_form}                
\end{align}

However, this approach is just the first order approximation of the standard WKB method. To obtain the QNMs of more complicated potentials, WKB method is a convenient procedure which offers good accuracy. This method was originally suggested in \cite{Schutz:1985zz}, and developed to the 3rd order beyond the eikonal approximation in \cite{Iyer:1986np}. It should also be noted that there has been considerable development in the accuracy in the Ferrari-Mashoon procedure and results up to 6th order in WKB are provided in \cite{Konoplya:2003ii} (see \cite{Konoplya:2010kv} for an usage of the 6th order WKB formula to the scattering problem). Below we provide the results of the 6th order WKB formula that will be exploited subsequently.
 
In WKB method one starts with the Schrodinger like wave equation
\begin{align}
\Psi''(x) + (\omega^2 - V(x))\Psi(x) = 0 
\end{align}
where the potential $V(x)$ approaches  a constant at $x\to\pm\infty$ and at some intermediate value $x_0$, it rises to a maximum. For the present problem we  identify $x\equiv r_{\star}$ and $\Psi(x)\equiv Q(r(r_{\star}))$. This problem is now analogous to the quantum mechanical scattering problem from the peak of a potential barrier, where the turning points divide the potential into three regions. The solutions in those regions are then matched at the boundaries to obtain the energy spectrum. However for the higher order WKB extension, it turns out that an explicit match of the interior solutions to WKB solutions in the exterior regions to the same order is not necessary (see \cite{Konoplya:2010kv} for details). The result with 6th order WKB formula then has the form
\begin{align}
\omega^2 = V_0 - i \sqrt{-V_2} \bigg(n+\frac1 2 \bigg) + \sum_{i=2}^6 A_i  \qquad n=0,1,2,...
\label{6wkb}
\end{align}
where $A_i$'s represent i-th order correction in the WKB formula \emph{e.g.},
\begin{align}
A_2 = ~&(-11 V_{3}^2 + 9 V_2 V_4 - 30 V_{3}^2 n + 18 V_2 V_4 n - 30 V_{3}^2 n^2 + 18 V_2 V_4 n^2)/(144 V_{2}^2) \\
\frac{i A_3}{\sqrt{-2 V_2}} = ~& (-155 V_{3}^4 + 342 V_2 V_{3}^2 V_4 - 63 V_{2}^2 V_{4}^2 - 156 V_{2}^2 V_3 V_5 + 36 V_{2}^3 V_6 - 545 V_{3}^4 n \nonumber \\ 
&  + 1134 V_2 V_{3}^2 V_4 n - 177 V_{2}^2 V_{4}^2 n - 480 V_{2}^2 V_3 V_5 n + 96 V_{2}^3 V_6 n - 705 V_{3}^4 n^2 + 
\nonumber \\
& 1350 V_2 V_{3}^2 V_4 n^2 - 153 V_{2}^2 V_{4}^2 n^2 - 504 V_{2}^2 V_3 V_5 n^2 + 72 V_{2}^3 V_6 n^2 - 470 V_{3}^4 n^3 
\nonumber \\
& + 900 V_2 V_{3}^2 V_4 n^3 - 102 V_{2}^2 V_{4}^2 n^3 - 336 V_{2}^2 V_3 V_5 n^3 + 48 V_{2}^3 V_6 n^3)/(6912 V_{2}^5) .
\end{align} 
Other correction terms $A_4,A_5, A_6$ are given in \cite{Konoplya:2003ii}. Here $V_0 (\tilde{r}_{\star})$ is the value of the effective potential in its maximum ($r=\tilde{r}_{\star})$ and $V_i(\tilde{r}_{\star})$, is the i-th derivative of $V$ with respect to tortoise coordinate in the maximum.


\section{Spacetime for a smeared (Gaussian) matter distribution}
\label{gwgauss}

In this section, we consider the metric of a spherically symmetric spacetime geometry with a Gaussian distributed matter source (this kind of matter distribution, motivated by Quantum Gravity perspective, has an astrophysical interest as well, as mentioned in Sec.\ref{introduc}). Our aim is to compute the QNM frequencies for this smeared system (to $6$'th order in WKB scheme). The first task is to find the form of the potential for odd parity perturbations of this background geometry.

\subsection{Perturbation of the spacetime }
\label{gwgauss1}

Let us now start with the QG-inspired spherically symmetric spacetime. The metric of such spacetime is given by \cite{Nicolini:2008aj}
\begin{align}
ds^2 = -\bigg(1-\frac{4M}{r\sqrt\pi}\gamma(3/2, {r^2}/{4\Theta})\bigg)dt^2+\bigg(1-\frac{4M}{r\sqrt\pi}\gamma(3/2, r^2/{4\Theta}) \bigg)^{-1}dr^2 + r^2 d\Omega^2
\label{qg_sphe}
\end{align} 
where, $d\Omega^2 = d\theta^2 + \sin^2\theta d\phi^2$, $\sqrt{\Theta}$ is some minimal length scale which removes the singularity of the usual Schwarzschild spacetime, and 
\begin{align}
\gamma(3/2, r^2/{4\Theta})=\int_{0}^{r^2/{4\Theta}}\sqrt t\,e^{t}\,dt 
\end{align}
is the lower incomplete Gamma function. If we now expand the incomplete Gamma function of eqn.~\eqref{qg_sphe} in the limit $r^2>>4\Theta$, the metric takes the following form \cite{Nicolini:2008aj}
\begin{align}
ds^2 = -\bigg(1- \frac{2M}{r} + \frac{2M}{\sqrt{\pi\Theta}}e^{-r^2/{4\Theta}}\bigg)dt^2+\bigg(1- \frac{2M}{r} + \frac{2M}{\sqrt{\pi\Theta}}e^{-r^2/{4\Theta}}\bigg)^{-1}dr^2 + r^2 d\Omega .
\label{mod_sph}
\end{align}


The perturbed Einstein equation in vacuum is then
\begin{align}
R_{\mu\nu} = \delta R_{\mu\nu} = 0 \qquad \text{as,}\qquad \bar{R}_{\mu\nu}=0 . 
\end{align}
This equation has ten components. It turns out that only three of them (corresponding to the components $\mathbf{\delta R_{r \phi}}, \mathbf{\delta R_{t \phi}}$ and $\mathbf{\delta R_{\theta \phi}}$) survive and they are respectively given below in explicit form

\begin{eqnarray} 
\bigg( 1-\frac{2M}{r}+\frac{2M}{\sqrt{\pi\Theta}}e^{-r^2/{4\Theta}} \bigg) \bigg[\partial^{2}_{rr}h_0  - \partial_{r}\partial_{t}h_1 
+ \frac{2}{r} \partial_{t}h_1 \bigg] - \frac{l(l+1)}{r^2}h_0  \nonumber \\
 + \frac{2}{r}\bigg(\frac{2M}{r^2}-\frac{Mr}{\Theta\sqrt{\pi\Theta}}e^{-r^2/{4\Theta}}\bigg)h_0 = 0 ,
\end{eqnarray}

\begin{eqnarray}
\bigg(1-\frac{2M}{r}+\frac{2M}{\sqrt{\pi\Theta}}e^{-r^2/{4\Theta}} \bigg)^{-1} \bigg[\partial^{2}_{tt}h_1 - \partial_{r}\partial_{t}h_0 
 + \frac{2}{r} \partial_{t}h_0 \bigg] +  \frac{(l+2)(l-1)}{r^2}h_1  = 0 ,
\end{eqnarray} 

\begin{eqnarray}
\partial_r \bigg[\bigg( 1-\frac{2M}{r}+\frac{2M}{\sqrt{\pi\Theta}}e^{-r^2/{4\Theta}} \bigg)h_1\bigg]  + 
\frac{\partial_t h_0}{\bigg( 1-\frac{2M}{r}  + \frac{2M}{\sqrt{\pi\Theta}}e^{-r^2/{4\Theta}} \bigg)}  = 0 
\label{eq_tehi}
\end{eqnarray}
where $h_0,h_1$ have been introduced earlier in Sec.~\ref{methqnms}(a).

Let us now introduce a $\Theta$-dependent generalized Regge-Wheeler variable $Q_{\Theta}$ as, 
\begin{align}
Q_{\Theta}(t,r) = \bigg(1- \frac{2M}{r} + \frac{2M}{\sqrt{\pi\Theta}}e^{-r^2/{4\Theta}}\bigg) \frac{h_1(t,r)}{r} .
\label{rw_va}
\end{align}
With the help of eqn.~\eqref{eq_tehi} we can eliminate $h_0(t,r)$ and thus the final equation for the axial perturbation assumes the following simple form,
\begin{align}
\frac{\partial^{2}}{\partial t^2}Q_{\Theta}(t,r) - \frac{\partial^{2}}{\partial r_{\Theta}^2}Q_{\Theta}(t,r) + V^{\Theta}_{\text{axial}}(r)Q_{\Theta}(t,r) = 0,
\label{rg_eq}
\end{align}
where $Q_{\Theta}(t,r)$ is given in eqn.~\eqref{rw_va} and the potential function is 
\begin{align}
V^{\Theta}_{\text{axial}}(r) = \bigg(1-\frac{2M}{r}+\frac{2M}{\sqrt{\pi\Theta}}e^{-r^2/{4\Theta}}\bigg)\bigg[\frac{l(l+1)}{r^2}-\frac{6M}{r^3}
&+ \frac{M}{\Theta\sqrt{\pi\Theta}}e^{-r^2/{4\Theta}} \nonumber \\
&+ \frac{4M}{\sqrt{\pi\Theta}\,r^2}e^{-r^2/{4\theta}}\bigg] \label{new_pot}.
\end{align}
We have defined the new co-ordinate $r_{\Theta}$, in analogy with the definition of eqn.~\eqref{star_old}, as
\begin{align}
\frac{dr_{\Theta}}{dr} = \frac{1}{(1- \frac{2M}{r} + \frac{2M}{\sqrt{\pi\Theta}}e^{-r^2/{4\Theta}})} .
\label{star_newr}
\end{align}
Likewise eqn.~\eqref{schr} we write $ Q_{\Theta}(t,r)$ as $Q_{\Theta}(t,r) \sim e^{i\omega^{\Theta} t} Q_{\Theta}(r)$. So, eqn.~\eqref{rg_eq} now becomes
\begin{align}
\frac{\partial^2 Q_{\Theta}(r)}{\partial r_{\Theta}^2} + \big[(\omega^{\Theta})^2 -V^{\Theta}_{\text{axial}}(r)\big]Q_{\Theta}(r) = 0 .
\label{rweq}
\end{align}

Thus, we have obtained the form of the potential for the odd parity perturbation of this spacetime. Here $(\omega^{\Theta})^2$ plays the role of energy whose numerical estimate is what we are interested in. 



\subsection{QNM due to smeared matter distribution}
\label{gwgauss2}

In this section we will evaluate the QNM frequencies for the potential given in eqn.~\eqref{new_pot}. Following the line of discussion made in Sec.~\ref{methqnms}, we will first determine the QNMs by employing the Ferrari-Mashoon method (\emph{i.e.} first order WKB) and subsequently we will also use the 6th order WKB formula for computing QNMs with much better precision.
 

To apply Ferrari-Mashoon method, we need to see whether this potential can also be mapped to the so-called P$\ddot{\text{o}}$schl-Teller potential of eqn.~\eqref{pt_pt}. Let us write the modified potential for axial perturbation after incorporating the $\Theta$-correction as,
\begin{align}
V^{\Theta}_{\text{axial}}(r) &= V_{\text{axial}} (r) + V_{\text{axial}}^{\text{extra}}(r) \label{fu_po} 
\end{align}
where $V_{\text{axial}}$ is given in eqn.~\eqref{sc_po} and 
\begin{align}
 V_{\text{axial}}^{\text{extra}}(r) = & \frac{2M}{\sqrt{\pi\Theta}}e^{-r^2/{4\Theta}}\bigg[ \frac{l(l+1)}{r^2}-\frac{6M}{r^3}
+\bigg( \frac{1}{2\Theta} + \frac{2}{r^2} \bigg) \bigg(1-\frac{2M}{r}\bigg) \bigg] .
\label{qg_part}
\end{align}
Just by inspection of eqn.~\eqref{qg_part} we can see that at large distance from the horizon at $r=2M$, the correction terms fall exponentially fast. Therefore, this effective potential is not expected to deviate much from $V_{PT}$. To see that, we find out the minimum of the $\Theta$-corrected effective potential in eqn.~\eqref{fu_po}. 

We assume that the extremum of the new potential $V^{\Theta}_{\text{axial}}(r)$ is perturbatively shifted to the point,
\begin{align}
r_{0}^{\Theta} \simeq ~~ r_0 + \frac{2M}{\sqrt{\pi\Theta}}\,e^{-r_{0}^2/{4\Theta}}\,r^{\prime}
\label{minima_new}
\end{align}
where $r^{\prime}$ is so far unknown and $r_0$ is the minimum of the effective potential of eqn.~\eqref{sc_po}, given by
\begin{align}
\frac{r_0}{M} = \frac{\sqrt{9(l^2+l+3)^2 -96 l(l+1)}}{2l(l+1)} + \frac{3(l^2 +l+3)}{2l(l+1)} .
\end{align}
Now taking the derivative of eqn.~\eqref{fu_po} and using eqn.~\eqref{minima_new}, we solve for $r^{\prime}$ to get the new extremum located at 
\begin{align}
r_{0}^{\Theta} &\simeq  \,\, r_0 + \frac{2M}{\sqrt{\pi\Theta}}\,e^{-r_{0}^2/{4\Theta}}\times\frac{1}{\Delta}\big[r_{0}^2 A + 
r_{0}^4 B \big]  \\
\text{where,}\quad A &= -2M\big[r_{0}^4 +12r_{0}^2\Theta +60\Theta^2\big], \nonumber\\
B &= \big[r_{0}^4 + 2(l^2 + l + 2)\Theta(r_{0}^2 + 4\Theta) \big], \nonumber \\
\Delta &=  24[ l(1 + l) r_{0}^2 - 4 (l^2 + l + 3) M r_0 + 40 M^2 ] \Theta^2 .  \nonumber            
\end{align}   

We note that for the $\Theta$-corrected metric of eqn.~\eqref{mod_sph}, the co-ordinate $r_{\Theta}$ is a function of the parameter $\Theta$.  Integrating eqn.~\eqref{star_newr} numerically, we found that the behaviour of this new coordinate $r_{\Theta}$ is not markedly different from $r_{\star}$, since $\frac{dr_{\star}}{dr}\sim\frac{dr_{\Theta}}{dr}$.
Therefore, the maximum of the potential $V^{\text{axial}}_0$ and the curvature parameter $\alpha^{\text{new}}$ are then given by,
$V^{\Theta}_0 = V_{\text{axial}}^{\Theta}(r) \big| _{r=r_{0}^{\Theta}}$ and 
$\alpha^{\Theta} =  \frac{1}{2V^{\Theta}_0}\,\frac{d^2 V_{\text{axial}}^{\Theta}(r)}{dr_{\Theta}^2}\bigg|_{r=r_{0}^{\Theta}}$
where  $d^2/dr_{\Theta}^2$ can be found by using the new transformation rule of eqn.~\eqref{star_newr}.

\begin{figure}[t!]
    \centering
 \includegraphics[width=8.1cm]{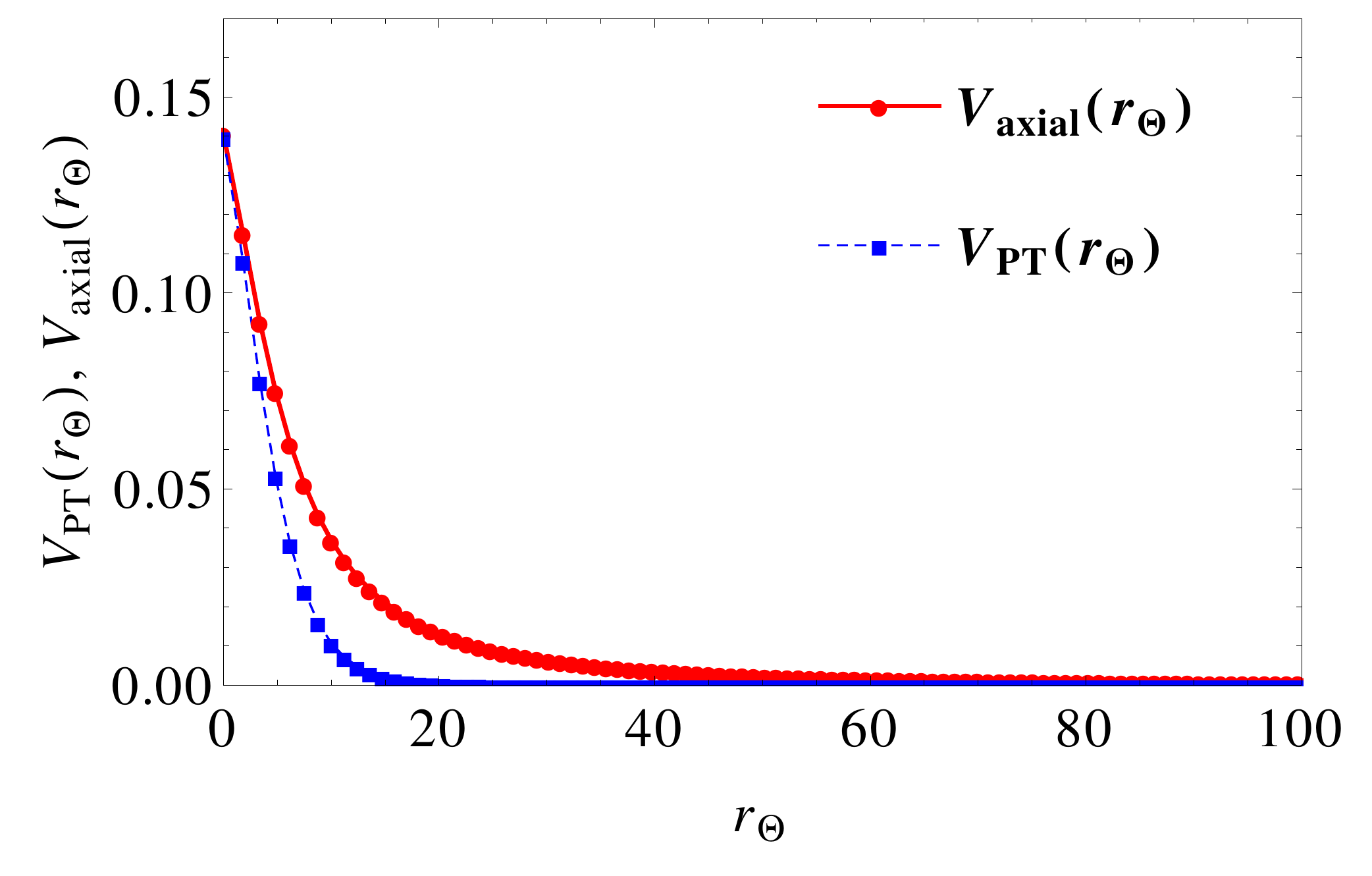}
 \includegraphics[width=8.1cm]{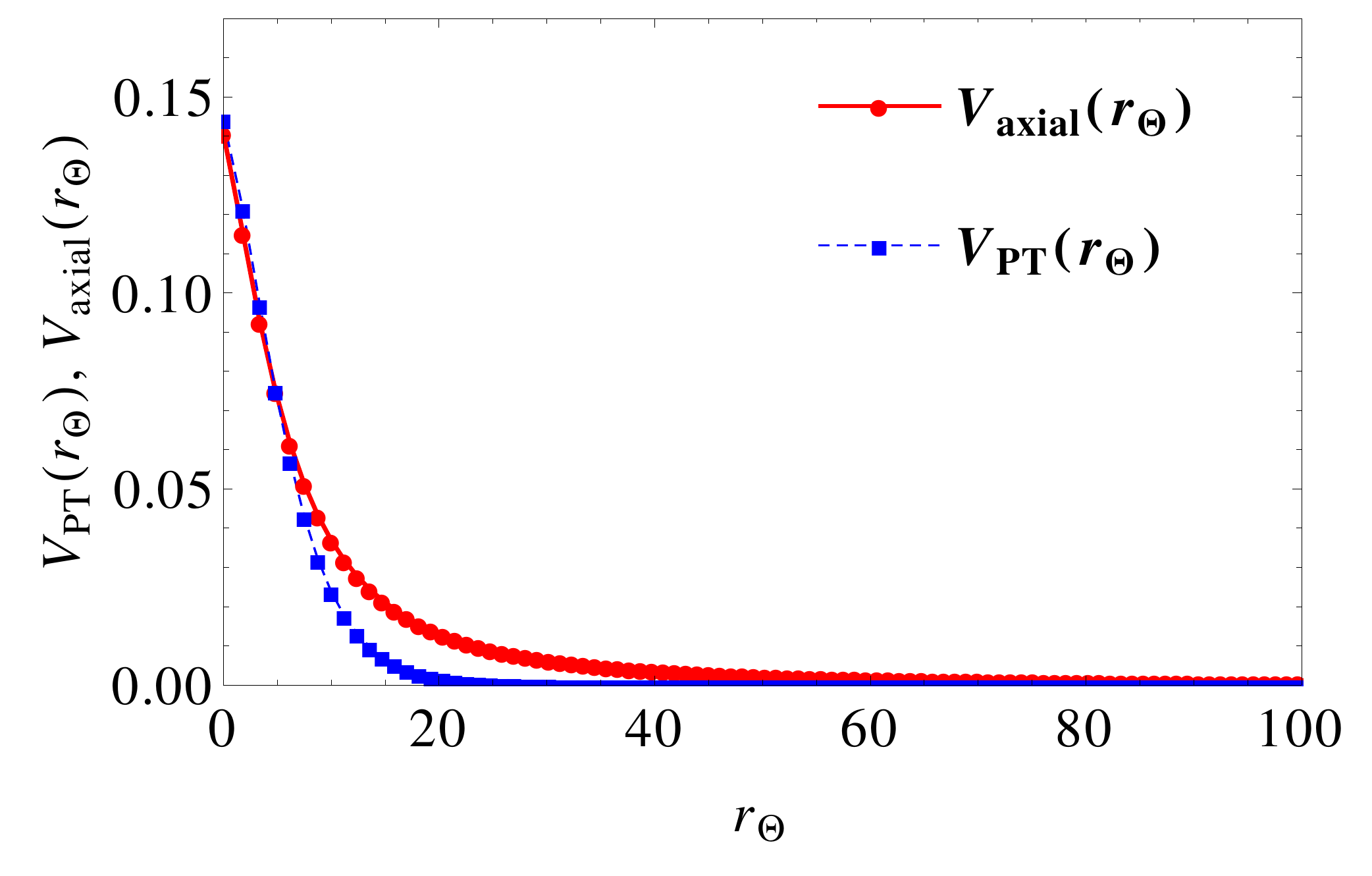}
             \caption{Plot showing the comparison of $V_{\text{axial}}$ (red) and $V_{PT}$ (dashed blue) as a function of the new coordinate $r_{\Theta}$ for $\Theta=0.3$, and $0.4$ respectively. The figure shows that the asymptotic behavior of both potentials are very close to each other.} 
\label{qnm_pot}
\end{figure}   

     
In Fig.~\ref{qnm_pot} we plot  variation of the P$\ddot{\text{o}}$schel-Teller potential  $V_{PT}$  and the potential for the axial perturbation $V_{\text{axial}}$  as a function of the co-ordinate $r_{\Theta}$. We have taken two different values of $\Theta$ to show that the form of the axial perturbation potential $V_{\text{axial}}^{\Theta}$ does not differ drastically from the form of $V_{PT}$ with respect to the new transformed co-ordinate $r_{\Theta}$. Since the QNMs of $V_{\text{axial}}^{\Theta}$ are related with the asymptotical behaviour of the potential, therefore one may legitimately use eqn.~\eqref{mash_form} to get the QNMs for this potential.  


As mentioned in Sec.~\ref{methqnms}, the semi-analytic method due to Ferrari-Mashoon is one of the easiest tool to estimate the QNMs. However, the  WKB treatment \cite{Schutz:1985zz} for computing QNMs offers much better accuracy. Also, in the context of QNM determination, the 3rd order WKB formula \cite{Iyer:1986np} was frequently used in the literature (see refs. \cite{Konoplya:2002ky,Kokkotas:1993ef,Andersson:1996xw,Onozawa:1995vu}).  Later it was shown that the WKB formula, when extended to the 6th order, gives the relative error which is about two orders less than that of the 3rd WKB order \cite{Konoplya:2004ip}. Therefore, in this work we will also compute the QNMs for the  potential $V_{\text{axial}}^{\Theta}$ using the 6th order WKB formula given by eqn.~\eqref{6wkb}. A previous study of QNMs for different perturbations of the $\Theta$-corrected  metric was done by Liang \cite{Liang:2018nmr,Liang:2018uyk} using 3rd order WKB formula.  
But we will shortly see that with the 6th order WKB formula, the associated QNMs will show non-trivial modifications in their numerical estimates. Finally, we will make a comparison of the results with various orders of the WKB method.

\subsection{Results of QNM} 
\label{gwgauss3}

\begin{figure}[t!]
    \centering
    \includegraphics[width=10.2cm]{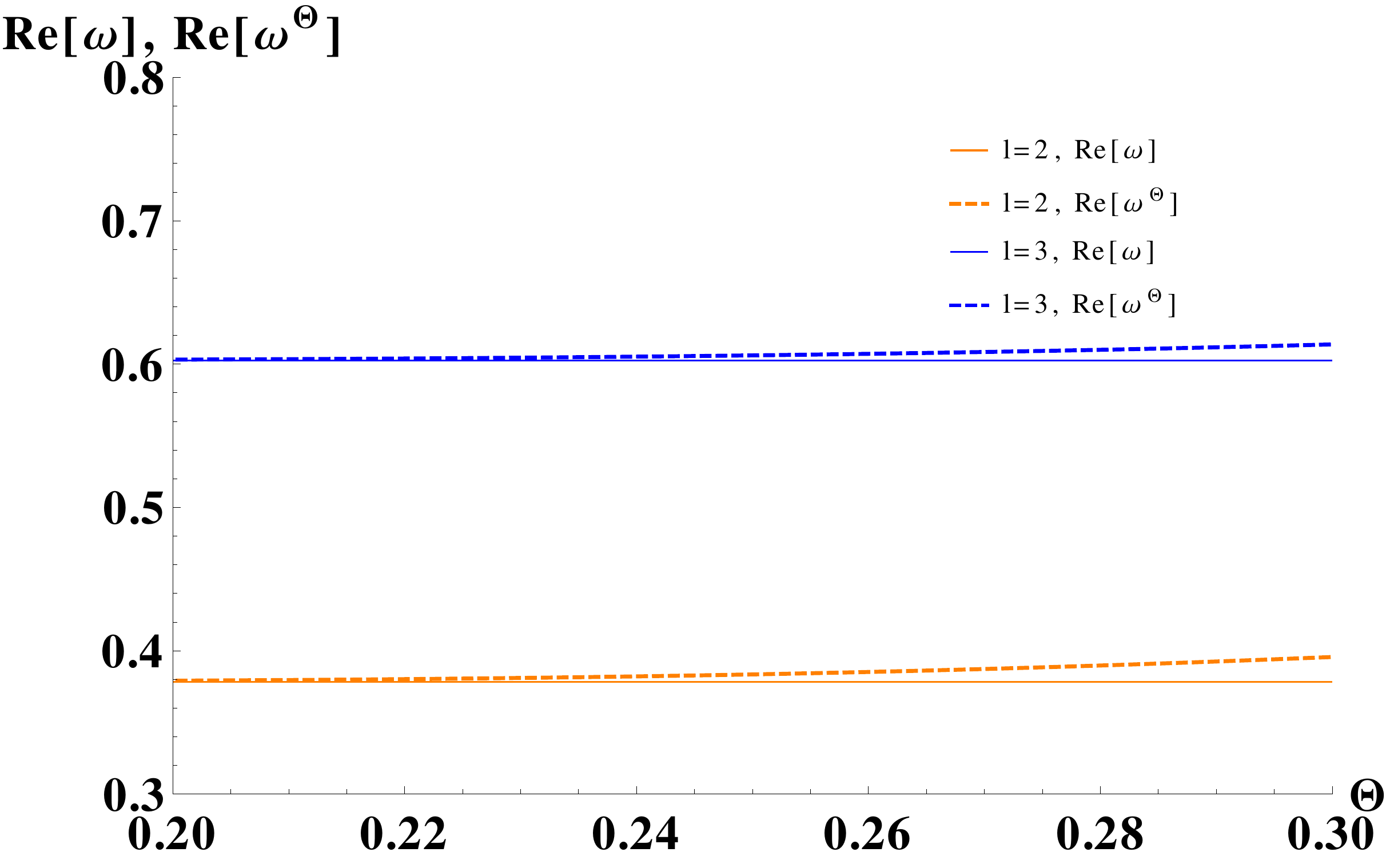}
             \caption{Plot showing the variation of the real part of QNM frequency $\omega^{\Theta}$, with increasing value 
of the parameter $\Theta$. For the above plot we have considered $l=2,~3$.  } 
\label{mfig}
\end{figure}

In fig.~\ref{mfig} we plot (colored dashed plots) the variation of Re$[\omega^{\Theta}]$ as a function of $\Theta$ using Ferrari-Mashoon method (\emph{i.e.} 1st order WKB). The associated frequencies Re$[\omega]$ (corresponding to the normal case with potential of eqn.~\eqref{sc_po}) in this graph are also shown by colored continuous plots. Since Re$[\omega]$ being independent of $\Theta$ for all values $l$, so they appear to be parallel straight lines for the same set of $l$ values. Here, the difference (Re$[\omega^{\Theta}]-$Re$[\omega]$) is found to increase for higher $l$ values (though shown for $l$ upto 2, but this is true for $l>2$ also, as explicitely checked by us). But this simple observation does not remain strictly valid when   the 6th order WKB formula eqn.~\eqref{6wkb} is exploited. Therefore, while eqn.~\eqref{mash_form} captures the relevant changes in QNMs as a first hint, it is better to rely on the 6th order WKB formula for numerical precision.

\begin{figure}[t!]
    \centering
    \includegraphics[width=10.2cm]{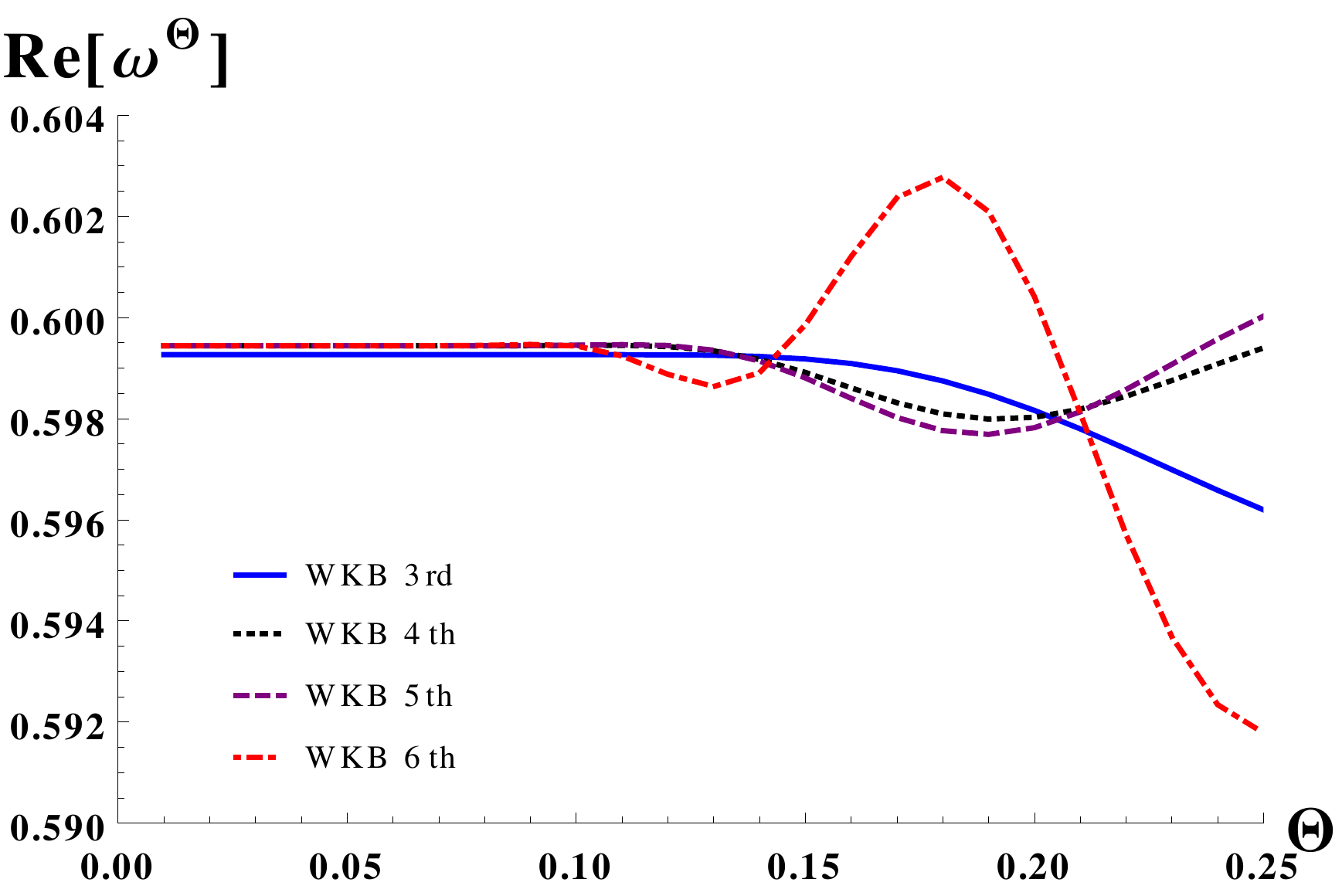}
             \caption{Plot showing the variation of the real part of QNM frequency $\omega^{\Theta}$, with increasing value 
of the parameter $\Theta$. For the above plot we have considered $l=3,~n=0$ and different orders of the WKB formula.  } 
\label{freq_plot}
\end{figure}

\begin{table}[h!]
\begin{center}
\begin{tabular}{|l|c|c|c|c|c|}
\hline
\hfill & \hfill & \hfill & \hfill & \hfill & \hfill \\
\bf l & \bf n & \bf Gravitational & $\Theta$ & \bf Gravitational & \bf Gravitational  \\
 &  & \bf modes   & & \bf modes &  \bf modes \\
 \cline{5-6}
&  & (Schwarzschild)  & & (Smeared matter) & (Smeared matter) \\
&  &  & & 3rd order WKB & 6th order WKB \\
\hfill & \hfill & \hfill & \hfill & \hfill & \hfill \\
\hline\hline
2 & 0 &  0.373616 - i\,0.0888891 & 0.1 & 0.373163 - i\,0.089221 & 0.374579 - i\,0.088672 \\
 & &   & 0.2 & 0.371753 - i\,0.089350 & 0.470351  - i\,0.044726   \\
 & &   & 0.3 & 0.360579 - i\,0.066717 & 0.309889  - i\,0.123928 \\
\cline{2-6} 
 & 1 & 0.346297 - i\,0.273480 &  0.1  & 0.346028 - i\,0.274930 & 0.352931  - i\,0.268630 \\
 &  &   & 0.2 & 0.339165 - i\,0.275108 & 0.918186  - i\,0.002729 \\
 &  &   & 0.3 & 0.234593 - i\,0.185924 & 0.235278  - i\,0.629412  \\
\hline 
3 & 0 &  0.599444 - i\,0.092703 & 0.1 & 0.599265 - i\,0.092732 & 0.599440  - i\,0.092735 \\
 & &   & 0.2 & 0.598163 - i\,0.091950 & 0.600413 - i\,0.090236 \\
 & &   & 0.3 & 0.594996 - i\,0.081382 & 0.596246 - i\,0.086833 \\
 \cline{2-6} 
 & 1 &  0.582642 - i\,0.281291 & 0.1 & 0.582362 - i\,0.281424 & 0.582510  - i\,0.281662 \\
 &  &   & 0.2 & 0.576212 - i\,0.278026 & 0.573718 - i\,0.276666  \\
 &  &   & 0.3 & 0.549907 - i\,0.239466 & 0.581571 - i\,0.261865 \\
\cline{2-6} 
 & 2 &  0.551594 - i\,0.479047 & 0.1 & 0.553235 - i\,0.476731 & 0.550833 - i\,0.481339 \\
 &  &   & 0.2 & 0.534658 - i\,0.469451 & 0.467130 - i\,0.566279 \\                             
 &  &   & 0.3 & 0.442211 - i\,0.393935 & 0.648012 - i\,0.366030 \\
\hline
\end{tabular} \vspace{0.3cm}
\end{center}
\caption{Comparison between the QNM frequencies for the gravitational perturbation of Schwarzschild spacetime 
and spherically symmetric spacetime 
 with smeared matter source.
}
\label{tab_gwv}
\end{table}
The following Table (Tab.\ref{tab_gwv}) shows the values of QNM frequencies for the odd parity gravitational perturbation of the $\Theta$-corrected space-time calculated using 6th order WKB formula.

\begin{figure}[t!]
    \centering
    \includegraphics[width=10.2cm]{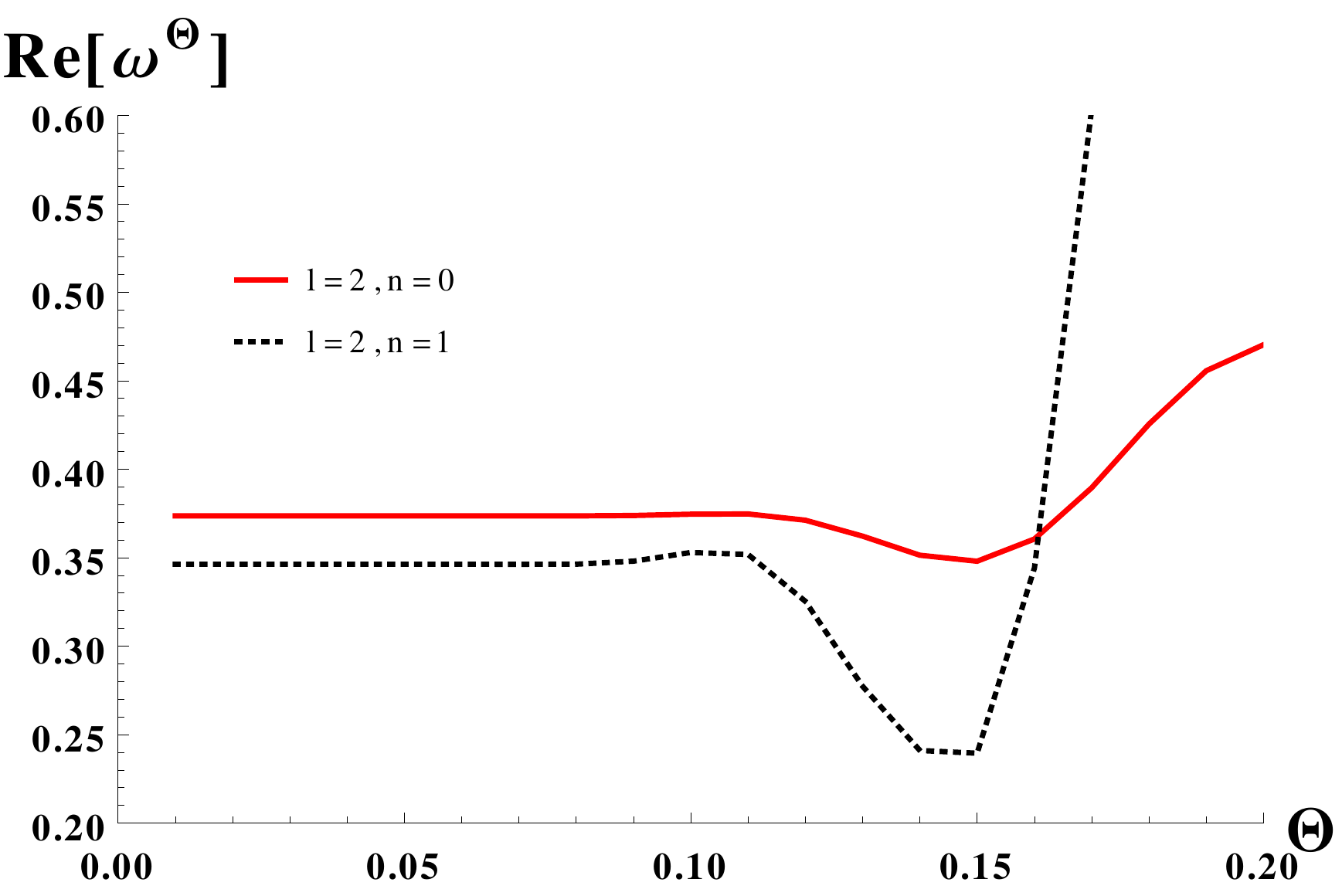}
             \caption{Plot showing the variation of Re$[\omega^{\Theta}]$ as a function of the parameter $\Theta$ for $l=2, n=0$ and $l=2, n=1$ using the 6th order WKB formula.  } 
\label{wkbm_plot}
\end{figure}
In fig.~\ref{freq_plot}, we plot the real part of QNM frequencies as a function of the parameter $\Theta$ for various orders of the WKB approximation. It is clear from the graph that why the higher order WKB gives significant alteration in the value of Re$[\omega^{\Theta}]$. For  $0.2>\Theta >0.01$, the 3rd order WKB result for the QNM frequency is nearly a constant. But this is not the correct picture if we go to 5th or 6th order approximations. In fact, Re$[\omega^{\Theta}]$ begins to decrese much earlier for $\Theta\gtrsim0.12$. This change is significant. Fig.~\ref{wkbm_plot} shows another plot for Re$[\omega^{\Theta}]$ for $l=2\,(n=0,1) $ modes computed using 6th order WKB formula. 
 
Let us compare our results which is valid up to 6th order in WKB with that of Liang \cite{Liang:2018nmr}  that is valid up to 3rd order in WKB. Liang has considered $\Theta $ to be lower than $0.25$ approximately. It is straightforward to check that for $\Theta > 0.25 $ there appears oscillations in the value of QNM frequency which seems to be spurious indicating that the 3rd order WKB perturbation scheme used by Liang is reliable up to $\Theta \sim 0.25$. On the other hand, from Fig.~\ref{freq_plot} it is clear that in 4th, 5th or 6th order in WKB such oscillations appear earlier at around $\Theta \sim 0.16$. Thus 6th order computational results restrict the value of $\Theta$ to lower values where the results are reliable. Naively it might seem that higher than 6th order results in WKB might restrict $\Theta$ further but as has been noted in \cite{Konoplya:2003ii}  orders of WKB higher than six are not feasible in the WKB framework.

\subsection{Observational aspects of $\Theta $-correction in QNM } 
\label{gwgauss4}

So far, we have  depicted the change in the QNM frequency spectrum when the source has smeared matter distribution. Here, we discuss the relevance of this result in the context of observational aspects. 
As an example, let us consider the fundamental GW mode (for $l=2, n=0$) of Schwarzschild geometry. The associated real part of the frequency $\omega_{re}= 0.373616$  from Tab.~\ref{tab_gwv} can be expressed in Hz unit as
\begin{align}
f= \frac{\omega_{re}}{2\pi M} \times \frac{c^3}{G} = \bigg(\frac{\omega_{re}}{2\pi}\times\frac{c^3}{GM_{\odot}}\bigg) 
\times\frac{M_{\odot}}{M} 
\label{conobs}
\end{align}
where $M_{\odot}$ is the solar mass. Using this formula with $M=1~M_{\odot}$, the frequency $f$ for the fundamental mode turns out to be $12$ kHz. 

Now there exist compact spherical star clusters (\emph{e.g} globular clusters) that approximately follow a Gaussian matter distribution. A typical order of magnitude estimate for the mass ($ \tilde M $) of such a  cluster is $10^5 M_{\odot}$. 
With $M=\tilde{M}$, it can be shown from eqn.~\eqref{conobs} that if the corresponding smeared distribution has a spread $\sqrt{\Theta}\sim 10^7 $ km (which matches with a $\Theta \sim 0.173$ within the range   $\Theta \sim 0.16 - 0.19$ of Tab.~\ref{tab_gwv}), then it yields a signal having frequency $f\sim13 $ kHz. As a first clue, this small change in frequency is significant to infer the nature of the source, that is to say, whether a GW detected with this frequency is associated to a  point mass or a diffused mass pattern. 



\section{Conclusion}
\label{conc}

In this work, we have studied the QNM frequency spectrum for the static spherically symmetric spacetime having a {\it {smeared}} (Gaussian type) matter distribution. This type of matter distribution, involving a length scale, can be motivated 
from astrophysical perspectives (in the context of star clusters). Hence our result can be relevant for those scenarios depending on a proper choice of the length of smearing  scale. Also, such a  length scale is crucial to identify the character of the source density.  As a demonstration with astrophysical objects, we found that the resulting frequency change due to smearing is of $\mathcal {O}(Hz)$ and hence within the current limits of the terrestrial GW detectors. Originally such metrics with smeared matter distribution was motivated by quantum gravity with the smearing length tentatively identified with Planck length. However, in that case, due to the smallness of Planck length scale such quantum gravity motivated corrections will be difficult to observe.


In summary, our analysis here focuses on the gravitational perturbations of a background geometry, that are odd multipoles under parity transformation. The gravitational perturbations do posses an even parity component as well. For the case of conventional spherically symmetric Schwarschild geometry with a delta-function source, there is a special property for the perturbation spectrum  that ensures that the QNM spectra for odd and even parity perturbations are equal. In technical terms, one says that  the QNM spectrum of odd parity perturbations is iso-spectral with the QNM spectrum of even parity perturbations \cite{Chandrasekhar:1985kt}.  However, it is not     clear whether  the same would  hold true for  spacetime with a smeared matter distribution, that has been studied here.  With  Gaussian distributed matter density, the potential for even-parity perturbation would have a new ($\Theta$-dependent) scale. This feature may restrict the validity of the iso-spectral character between perturbations of opposite parity. This requires  an explicit computation of the QNM spectrum of even-parity perturbations that we have left  as a  future work. 

{\bf{Acknowledgement:}} It is a pleasure to thank Professor Roman Konoplya for helpful suggestions and informing us about the relevant references for this work and specially  for sending us the necessary code for computation. Also we thank the referee for constructive comments.


\end{document}